\RequirePackage{fix-cm}
\documentclass[twocolumn,epjc3]{svjour3}

\RequirePackage{color,graphicx,url}

\journalname{Eur. Phys. J. A}

\usepackage{amsfonts,amsmath,amssymb,bm,xspace}

\graphicspath{{./figures/}}

\usepackage[colorlinks=true]{hyperref}  

\graphicspath{{figures/}} 
\usepackage{multirow}
\usepackage{subcaption}
\usepackage{academicons} 
\newcommand{\orcid}[1]{\href{https://orcid.org/#1}{\textcolor[HTML]{A6CE39}{\aiOrcid}}}
\usepackage{xcolor}

\usepackage{bm}
\usepackage{enumitem}

\usepackage{nccmath}
\usepackage{microtype}
\usepackage{xfrac}
\usepackage{soul}

\hypersetup{%
    pdfsubject=Paper,
    pdfkeywords={nuclear physics} {MBPT} {chiral EFT} {asymmetric nuclear matter}
    unicode=true,
    breaklinks=true,
     colorlinks   = true,
     linkcolor = blue,
     citecolor = blue,
     menucolor = blue,
     urlcolor = blue
}
\usepackage{cuted}
\newenvironment{widetext}{\begin{strip}}{\end{strip}}
\newcommand{\sat}{\mathrm{sat}}

\newcommand{\ri}{\mathrm{i}}

\newcommand{\tr}{\mathrm{tr}}

\begin{document}
\title{Internal Nucleon Properties and the Equation of State of Ultra-Dense Matter}
\author{B. K. Pradhan\thanksref{e1,addr1}\href{https://orcid.org/0000-0002-2526-1421}{\includegraphics[scale=0.3]{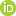}} 
\and
J. Margueron\thanksref{addr2}\href{https://orcid.org/0000-0001-8743-3092}{\includegraphics[scale=0.3]{orcid.png}}
\and
G. Chanfray\thanksref{addr1}\href{https://orcid.org/0000-0002-3593-9507}{\includegraphics[scale=0.3]{orcid.png}}
\and
H. Hansen\thanksref{addr1}\href{https://orcid.org/0000-0001-8879-3612}{\includegraphics[scale=0.3]{orcid.png}}
}
\institute{Institut de Physique des 2 infinis de Lyon, CNRS/IN2P3, Universit\'e de Lyon, Universit\'e Claude Bernard Lyon 1, F-69622 Villeurbanne Cedex, France \label{addr1}
\and
International Research Laboratory on Nuclear Physics and Astrophysics, Michigan State University and CNRS, East Lansing, MI 48824, USA\label{addr2}
}

\thankstext{e1}{b-k.pradhan@ip2i.in2p3.fr}

\date{Draft version:\today, Received: date / Accepted: date}

\maketitle

\begin{abstract}
The question of the link between the nucleon properties and the equation of state of ultra-dense matter, both being composed of quarks and gluons, is investigated by adopting a soliton model to describe internal nucleon properties and a direct mapping between nucleon and quark matter to obtain the equation of state of ultra-dense matter. We compare $\pi+\omega$ and $\pi+\omega+\rho$ soliton models and analyze the presence of a pressure singularity at the nucleon center due to the $\rho$ field. This singularity impacts the equation of state in ultra-dense matter. Only the pressure derived from the canonical energy-momentum tensor has a singularity, while the Belinfante pseudo-gauge transformation of the energy-momentum tensor generates an internal pressure without a singularity. We then investigate predictions for the equation of state of ultra-dense matter based on different prescriptions for the nucleon pressure. It is shown that none of the prescriptions agree with the thermodynamical pressure defined in uniform matter. The mismatch between any equation of state and the thermodynamical one questions the simplicity of the direct mapping between finite-size nucleon properties and uniform matter as well as the role of the nucleon surface in the internal energy density and the pressure.
\end{abstract}
\section{Introduction}

In the nucleon's center, the particle density far exceeds the nuclear saturation density ($n_{\rm sat}\approx 0.16$~fm$^{-3}$), motivating recent attempts to connect the properties of ultra-dense matter, where the density is typically 5-8$n_\sat$, with the properties of the nucleon, both of which being composed of quarks and gluons~\cite{Lorce2018,Fukushima2020,Pradhan2025}. The interest in this connection is that experimental data associated with the description of the nucleon, such as those available in Refs.~\cite{Burkert2018,PDG2020}, could be employed to access the properties of cold quark and gluon matter~\cite{Shanahan2019}. It is therefore very tempting to consider the nucleon as a piece of cold quark and gluon matter and extract its bulk properties, especially since very few experimental constraints exist for ultra-dense quark matter, except the one from perturbative QCD (pQCD) at large particle density, i.e. about 40-50$n_\sat$~\cite{Blaizot2003,Kajantie2003,Kurkela2010,Collins2011}. Note that a connection between the properties of matter at densities 5-10$n_\sat$ and pQCD was recently suggested \cite{Shirke2022,Annala2019,Annala2022,Annala2023,Komoltsev2025,Gorda2023,Fan2024} but the constraining power of this connection with respect to other constraints, e.g., astrophysical data, is still under discussion \cite{Somasundaram:2023b,Komoltsev2023}.

To connect nucleons and ultra-dense matter, one may first consider a nucleon model that provides quantities such as particle density, energy density, and pressure, which can be compared with experimental data. Then, in a second step, one may define how quantities well defined in the nucleon model are mapped into uniform matter and employed in predicting the ultra-dense matter equation of state (EoS). In recent works, a direct mapping between these two systems was suggested~\cite{Lorce2018,Fukushima2020}. This direct connection has been explored based on the soliton model for the nucleon~\cite{Fukushima2020,Pradhan2025}.

In this paper, we consider two soliton models: a simple $\pi+\omega$ soliton model, stabilized by the $\omega$ field, and a more complete $\pi+\omega+\rho$ model, which includes the isovector field $\rho$, important for addressing neutron star (NS) properties where matter is dominantly composed of neutrons. The latter is the same model as in Refs.~\cite{Fukushima2020,Pradhan2025} and is the one finally selected in our study. In NS, the pressure entering into the Tolman-Oppenheimer-Volkoff (TOV) equations is the hydrodynamical pressure~\cite{Tolman1939,Oppenheimer1939}, which can be obtained from the mechanical pressure, derived in the soliton model, or from the thermodynamic expression. An important question is how to connect these two systems, one being finite-size (the nucleon) and the other being infinite (ultra-dense matter). The direct mapping suggested in Refs.~\cite{Lorce2018,Fukushima2020,Pradhan2025} neglects the possible surface contribution. If the conditions to perform this direct mapping were satisfied (no or negligible surface contribution), then the mechanical and thermodynamic pressure would coincide in uniform matter. We, however, show in this paper that it is not the case.

The difference between these pressures is probably related to the direct mapping itself, and thus, to the treatment of the surface term in finite-size systems. Note that there are examples where, even in infinite systems, the mechanical and the thermodynamic pressure differ; see, for instance, Ref.~\cite{MdHijes:2026}. When it occurs, it reflects that there are still surface contributions in the infinite system, due, for instance, to the non-uniformity of matter. So, the difference between the mechanical and thermodynamic pressure, when it occurs, reflects the fact that the surface term is not well treated. The scope of this paper is to analyze the properties of ultra-dense matter and quantitatively estimate the difference between induced EoS.

In addition, two expressions for the mechanical pressure can be obtained depending on whether it is derived from the canonical energy-momentum tensor (EMT) or from the Belinfante EMT, which are connected by a pseudo-gauge transformation; see Ref.~\cite{Fukushima2026} and references therein. When the $\rho$ meson is included, the canonical EMT induces a singularity in the mechanical pressure at the center of the nucleon~\cite{Fukushima2026}, while the pressure from the Belinfante EMT remains regular. Note that this singularity does not impact the size of the soliton, as we will analyze it in this paper. We also investigate a heuristic approach (mixed model), which is not based on a theoretical derivation but has been employed in recent studies~\cite{Fukushima2020,Pradhan2025}: the energy density is obtained from the canonical EMT while the pressure is that derived from the Belinfante EMT. 
Since the canonical and Belinfante prescriptions for the EMT are related by a pseudo-gauge transformation, they are not expected to impact the EoS of ultra-dense matter~\cite{Fukushima2026}. The fact that there is an impact may pinpoint a caveat in the calculation of the EMT. 

The surface contribution to the energy and pressure in finite systems, such as the nucleon, is an important quantity to consider before mapping the thermodynamics in the nucleon and in uniform matter. This contribution originates from the density change marking the frontier between the inside and the outside regions of finite systems. It receives contributions from the finite range of the strong interaction, also called the gradient term in density functional theories, as well as from the variation of the density at the surface. In nuclei, for instance, the surface energy receives contributions from both the finite range of the nuclear interaction and the gradient of the density, in almost equal parts~\cite{RingSchuck1980,Bender2003,Reinhard2006}. It is often defined as the difference between the properties of the finite-size system and its uniform matter analog. In the case of the soliton model, the uniform analog does not exist, and thus, it is difficult to define the surface contribution to the energy or pressure. It is possible, instead, to compare the mechanical and thermodynamic definitions of the pressure and employ them in the prediction of the ultra-dense matter EoS. Since these two definitions of the pressure must be equal in ultra-dense matter, the observed differences reflect the incompleteness of the soliton model for describing ultra-dense matter, in particular, the impossibility of defining the surface contribution to the nucleon pressure.

The paper is organized as follows: The formalism for the $\pi+\omega$ and $\pi+\omega+\rho$ soliton models, as well as the canonical and Belinfante energy-momentum tensors, are briefly described in Sec.~\ref{sec:Formalism}, and the results, including a detailed comparison between these different expressions and the thermodynamic pressure for the equation of state in ultra-dense matter, are presented in Sec.~\ref{sec:qm}. We then present our conclusions and outlook in Sec.~\ref{sec:conclusions}.

\section{The soliton model}
\label{sec:Formalism}

The soliton model can be derived from the 2-flavor NJL model of light quarks using the bosonization approach and a gradient expansion, as detailed in Refs.~\cite{Zuckert1995b,Pradhan2025}. 
Note that the Skyrmion (explicitly the NJL-chiral soliton) description of the nucleon is stabilized by the vector $\omega$ and iso-vector $\rho$ mesons. 

\subsection{Lagrangian}

The soliton model is described by the Lagrangian involving a pion field, embedded in the unitary operator $U=\xi^2$, and a vector iso-scalar ($\omega_{\mu}$) and a vector iso-vector ($\bm{\rho}_{\mu}$) fields. 
The Lagrangian density is given as:
\begin{eqnarray}
\label{eq:Lagrangian}
\mathcal{L} &=& \frac{1}{4}f_\pi^2\tr(\partial_\mu U \partial^\mu U^\dagger) - \frac{1}{2}f_\pi^2\tr(D_\mu\xi\cdot\xi^\dagger+D_\mu\xi^\dagger\cdot\xi)^2 \nonumber \\
&&\hspace{-1.0cm}- \frac{1}{2g^2}\tr(F_{\mu\nu}F^{\mu\nu}) + \frac{3}{2} g\omega_\mu B^\mu + \frac{1}{2}m_\pi^2f_\pi^2\tr(U + U^\dagger - 2)\, , \nonumber \\
\end{eqnarray}
where $B^{\mu}$ is the baryon current, $m_\pi$ is the mass of the $\pi$ meson, $f_\pi$ is the $\pi$ decay constant, and $g$ is the model parameter. In the Lagrangian density, $D_\mu$ is the covariant derivative and $F_{\mu\nu}$ is the field strength tensor for the vector meson fields, given respectively as:
\begin{align}
D_\mu &= \partial_\mu - \ri V_\mu\,,\\
F_{\mu\nu} &= \partial_\mu V_\nu - \partial_\nu V_\mu - \ri[V_\mu, V_\nu]\,.
\end{align}
In the above expressions, $V_\mu$ represents the vector meson matrix in flavor space defined by
\begin{equation}
V_\mu = \frac{g}{2} (\bm{\tau}\cdot\bm{\rho}_\mu + \omega_\mu) \, . 
\end{equation}

It is also possible to consider a soliton model without the $\rho$ meson: the $\pi+\omega$ soliton model was introduced in Ref.~\cite{ADKINS1984251}. This model is stabilized by the vector $\omega$ field only. In this case, the vector meson matrix in flavor space is defined by
\begin{equation}
V_\mu^{\pi+\omega} = \frac{g}{2} \omega_\mu \, . \label{eq:Vmu_pi_omega}
\end{equation}
In the following, we compare the results obtained for the soliton model with $\pi+\omega+\rho$ (hereafter named the soliton model) with the $\pi+\omega$ model. We refer the reader to Ref.~\cite{Meissner1986} for details on the equations of motion for the meson profiles and the EMT of the $\pi+\omega$ soliton.
 
\subsection{Profiles for the fields, density, energy-density and pressure}

Considering the soliton model to describe the properties of the nucleon, it is possible to define the energy density and the mechanical pressure profiles, $\varepsilon(r)$ and $p_\text{mec}(r)$~\cite{Fukushima2020}. In Ref.~\cite{Fukushima2026}, a two-flavor Skyrme model including vector mesons has been employed to study the spatial distribution of pressure, energy density, and shear forces. For a given Lagrangian $\mathcal{L}$, see Eq.~\eqref{eq:Lagrangian}, for instance, it was found that the EMT becomes pseudo-gauge dependent, creating an ambiguity between two definitions: the canonical EMT,
\begin{equation}
T_\text{can}^{\mu\nu} = \sum_{\phi}\frac{\partial\mathcal{L}}{\partial(\partial_\mu\phi)}\partial^\nu\phi - g^{\mu\nu}\mathcal{L}\,,
\label{eq:Tcan}
\end{equation}
and the Belinfante one,
\begin{equation}
T_\text{Bel}^{\mu\nu} = T_\text{can}^{\mu\nu} + \frac{1}{2}\partial_\lambda(S^{\lambda\mu\nu} - S^{\mu\lambda\nu} + S^{\nu\mu\lambda})\,,
\end{equation}
with
\begin{equation}
S^{\lambda\mu\nu}
= -\frac{2}{g^2}\,\mathrm{tr} \left(
F^{\lambda\mu} V^\nu - F^{\lambda\nu} V^\mu
\right)
\end{equation}
the spin current operator, and with $g$, $F^{\mu\nu}$ and $V^\mu$ defined in the following. These two definitions have quantitative consequences for the mechanical pressure profile inside a nucleon, producing a singularity at the center for the canonical EMT that the Belinfante EMT does not have. The pseudo-gauge invariance of the EMT has implications for the ultra-dense matter EoS if the local-space properties of the EMT are employed to determine the EoS. The EoS is, however, an observable, independent of the pseudo-gauge transformation of the EMT. In this paper, we do not address this question, but we rather investigate the consequences of the two definitions of the EMT in terms of uncertainties.

We consider the hedgehog ansatz for the pion field, with corresponding configurations for the vector mesons, specifically the Wu-Yang-’t Hooft-Polyakov ansatz for the $\rho$ meson and a time-component-only form for the $\omega$ meson due to the absence of spatial source terms in static solutions:
\begin{align}
U(\mathbf{r}) &= e^{i \boldsymbol{\tau} \cdot \hat{\mathbf{r}} F(r)}, \\
\rho^{i,a}(\mathbf{r}) &= \epsilon^{ika} \hat{\mathbf{r}}^k \frac{-G(r)}{g r}, \\
\omega^\mu(\mathbf{r}) &= \delta^{\mu 0} \omega(r),
\end{align}
From the conditions to minimize the energy, we get a set of equations of motion as follows:
\begin{align}
F^{\prime\prime} &= - \frac{2}{r}F^\prime - \frac{3g}{4\pi^2f_\pi^2r^2}\omega^\prime\sin^2F \notag \\
&\qquad + \frac{1}{r^2}[4(G + 1)\sin F - \sin2F] + m_\pi^2\sin F
\,, \label{eq:F_sol}\\
G^{\prime\prime} &= \frac{1}{r^2}G(G + 1)(G + 2) + 2g^2f_\pi^2(G + 1 - \cos F)\,, \label{eq:G_sol}\\
\omega^{\prime\prime} &= -\frac{2}{r}\omega^\prime + 2g^2f_\pi^2\omega - \frac{3g}{4\pi^2r^2}F^\prime \sin^2 F\, \label{eq:omega_sol}.
\end{align}
Finally, we consider the following boundary conditions for an isolated nucleon, and for the $\pi$ field $F$ and for the vector fields $G$ and $\omega$:
\begin{align}
&F(0) = \pi\,, && \hspace*{-3em} F(\infty) = 0\,,\\
&G(0) = -2\,, && \hspace*{-3em} G(\infty) = 0\,,\\
&\omega^\prime(0) = 0\,, && \hspace*{-3em} \omega(\infty) = 0\,.
\end{align}

\begin{figure}[t]
\centering
\includegraphics[width=\linewidth]{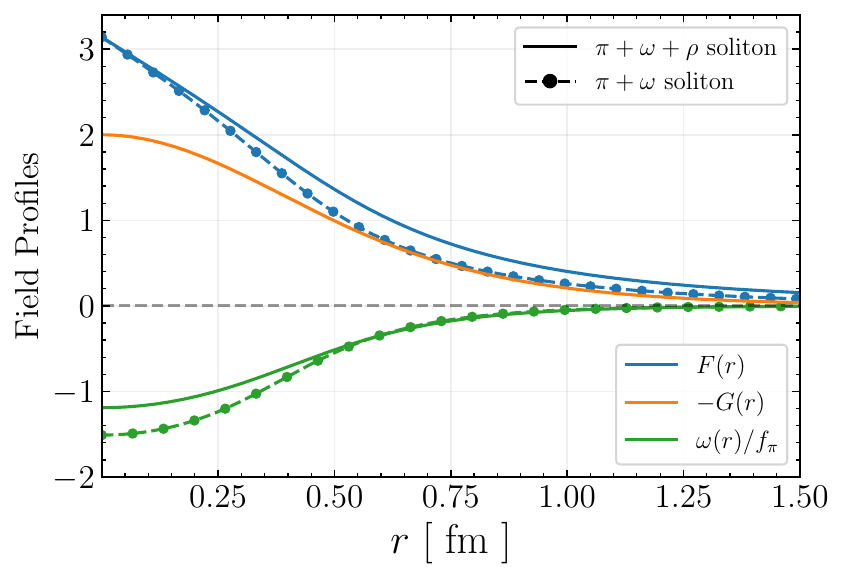}
\caption{Field profiles for $F$, $-G$ and $\omega/f_\pi$ obtained by solving Eqs.~\eqref{eq:F_sol}-\eqref{eq:omega_sol} (solid lines). For comparison, the field profiles corresponding to the $\pi+\omega$ soliton are also displayed as dashed lines with markers. Note that no profile for $G$ exists in the $\pi+\omega$ soliton. See text for more details.}

\label{fig:field_profiles}
\end{figure}

In the present work, we adopt the NJL-FKW parameter set~\cite{Pradhan2025}, where the meson masses $m_{\omega} = 783$~MeV and $m_{\pi} = 139$~MeV, the vector coupling constant is fixed at $g = 6$, and the pion decay constant is set to $f_{\pi} = 92.277$~MeV to satisfy the Kawarabayashi-Suzuki and Fayyazuddin-Riazuddin (KSFR) relation: $m_{\omega}^{2} = 2 g^{2} f_{\pi}^{2}$.

The field profiles obtained by solving the set of Eqs.~\eqref{eq:F_sol}-\eqref{eq:omega_sol} along with the aforementioned boundary conditions are displayed in Fig.~\ref{fig:field_profiles}. Note that the $F$ and $\omega$ field profiles are qualitatively similar for the $\pi+\omega$ and $\pi+\omega+\rho$ soliton models. There is, however, no $G$ field in the $\pi+\omega$ model.

The behavior of the fields close to $r\to 0$ can be analyzed. Considering the boundary conditions near the origin, we have:
\begin{eqnarray}
F(r) &\simeq& \pi - a r + \mathcal{O}(r^{2})\, , \label{eq:F_small_r} \\
G(r) &\simeq& -2 + b r^{2} + \mathcal{O}(r^{4})\, ,\label{eq:G_small_r} \\
\omega(r)/f_\pi &\simeq& c + d r^2 + \mathcal{O}(r^{3})\, \label{eq:omega_small_r}\, .
\end{eqnarray}
with $a\approx 3.39$~fm$^{-1}$, $b\approx5.95$~fm$^{-2}$, $c\approx -1.19$ and $d\approx 3.23$~fm$^{-2}$.

The baryon charge density $n_B(r)$ and the iso-scalar charge density $n_S(r)$ of the soliton can be obtained as;
\begin{align}
n_B(r) &= -\frac{1}{2 \pi^2 r^2} \, F'(r) \sin^2 F(r)\, \label{eq:rhob},\\  
n_S(r) &= -2 C gf_{\pi}^2 \, \omega(r) \label{eq:rhos}\, .
\end{align}
The boundary condition imposed for $F(r)$ ensures that the quantized baryon charge $B = \int_0^\infty\!d^3r\, n_B(r)=1$, representing the baryon number $B=1$ soliton. The constant $C$ in Eq.~\eqref{eq:rhos} is adjusted to normalize the iso-scalar charge density ($\int_0^\infty \!d^3r\, n_S(r)=1$). 

\begin{figure}[t]
\centering
\includegraphics[width=\linewidth]{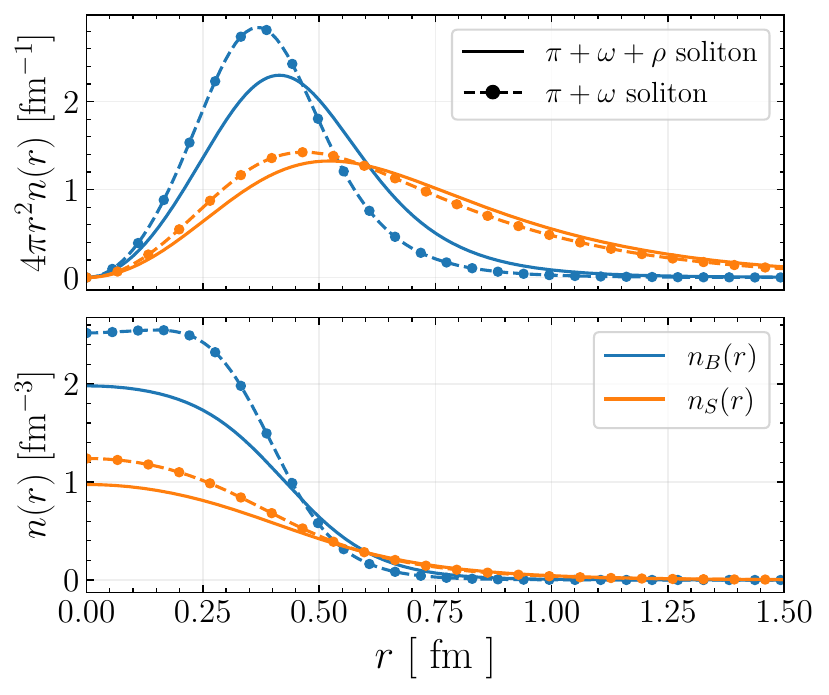}
\caption{Baryon (blue lines) and scalar (orange lines) densities are shown in the bottom panel, while the corresponding particle distributions are displayed in the top panel, as functions of the radial coordinate $r$. Solid lines represent the results for the $\pi+\omega+\rho$ soliton, whereas dashed lines correspond to the $\pi+\omega$ soliton.}
\label{fig:baryon_density_vsr}
\end{figure}

The baryon density $n_{B}(r)$ and isoscalar density $n_{S}(r)$ defined in Eqs.~\eqref{eq:rhob}-\eqref{eq:rhos} are represented in Fig.~\ref{fig:baryon_density_vsr}, for the same parameter set generating the field profiles shown in Fig.~\ref{fig:field_profiles}. The isoscalar charge density depends only on the $\omega$ field. Since the field $\omega(r)$ obtained from both the $\pi+\omega$ and $\pi+\omega+\rho$ models remains in close agreement throughout most of the soliton region, except at the nucleon center where small differences appear, the resulting isoscalar charge density distribution is also in good qualitative agreement between the $\pi+\omega$ and $\pi+\omega+\rho$ models. Furthermore, although this is not clearly visible in Fig.~\ref{fig:baryon_density_vsr}, the baryon density distributions derived from the $\pi+\omega$ and $\pi+\omega+\rho$ soliton models exhibit similar qualitative behavior.

The baryon density is associated with the distribution of matter and will be employed to map the EoS in the following. The scalar density is, however, more extended. Asymptotically, it drops to zero less fast than the baryon density, creating an intermediate region where the baryon density is negligible while the scalar density is still large. This region is sometimes referred to as a cloud or a halo~\cite{Fukushima2020}: In this core-and-cloud picture of the nucleon, the hard core, which hosts the valence quarks and carries the baryon density, remains comparatively robust against moderate changes in density until neighboring compact cores approach each other and begin to overlap. Well before such overlap occurs, the mesonic tails, associated with the surrounding cloud or halo, generate meson-exchange interactions. With increasing density, quark-antiquark pairs gradually delocalize, enhancing their mobility across multiple nucleons (it can be loosely pictured as the overlap of the cloud or halo region) in a percolation-like process. This regime is commonly referred to as ``soft'' deconfinement. As the density of nuclear or neutron star matter increases and reaches a critical value, the hard cores of neighboring nucleons begin to overlap, and quark matter is realized unambiguously. This transition is named ``hard deconfinement'' and may be identified with the deconfinement of valence quarks~\cite{Fukushima2020}.


From the field profiles of $F(r)$, $G(r)$, and $\omega(r)$, it is possible to determine the canonical energy density $\varepsilon_\text{can}(r)$, canonical mechanical pressure $p_\text{can}^\text{mec}(r)$  and canonical shear force $s_{\rm can}(r)$ as~\cite{Fukushima2026},

\begin{eqnarray}
\varepsilon_\text{can}(r) &=& \frac{1}{2}f_\pi^2\left({F^\prime}^2 + 2\frac{\sin^2 F}{r^2}\right) + f_\pi^2m_\pi^2(1 - \cos F) \nonumber \\
&\quad& + \frac{2}{r^2}f_\pi^2(G + 1 - \cos F)^2 - f_\pi^2g^2\omega^2 \nonumber \\ 
&\qquad& + \frac{1}{2g^2r^4}\left[2r^2{G^\prime}^2 + G^2(G + 2)^2\right] - \frac{1}{2} {\omega^\prime}^2 \nonumber \\
&\quad& + \frac{3g\omega}{4\pi^2r^2}F^\prime\sin^2 F\,, \label{eq:ecan} 
\end{eqnarray}
\begin{eqnarray}
p_\text{can}^\text{mec}(r)&=&- \frac{1}{6}f_\pi^2\left({F^\prime}^2 + 2\frac{\sin^2 F}{r^2}\right) - f_\pi^2m_\pi^2(1 - \cos F) \nonumber \\
&\quad& - \frac{2}{3r^2}f_\pi^2(1 - \cos F)(G + 1 - \cos F) \nonumber \\
&\quad& - \frac{2}{r^2}f_\pi^2G(G + 1 - \cos F)+ f_\pi^2g^2\omega^2 \nonumber \\
&\quad& - \frac{1}{6g^2r^4}\Bigl[2r^2{G^\prime}^2 + 3G^3(G + 2) + 2G^2(G + 2)\Bigr] \nonumber \\
&\quad& + \frac{1}{6}{\omega^\prime}^2\,, \label{eq:pcan}
\end{eqnarray}
\begin{eqnarray} 
s_\text{can}(r) &=&f_\pi^2\left({F^\prime}^2 - \frac{\sin^2 F}{r^2}\right)\nonumber \\
&\quad& - \frac{2}{r^2}f_\pi^2(1 - \cos F)(G + 1 - \cos F) \nonumber \\
&\quad& + \frac{1}{g^2r^4}\Bigl[2r^2{G^\prime}^2 - G^2(G + 2) - 3rGG^\prime\Bigr]\nonumber\\
&\quad&- {\omega^\prime}^2\, .\label{eq:shear_can}
\end{eqnarray}
As pointed out in Ref.~\cite{Fukushima2026} and discussed in the introduction, the nucleon's EMT can be modified through a pseudo-gauge transformation. The singularity in the canonical mechanical pressure can be removed by using the Belinfante EMT. For the ansatz considered for meson fields, one finds that the Belinfante energy density differs from the canonical one only through contributions from the $\omega$ meson (as non-zero for the time component). In contrast, the Belinfante pressure and shear force receive additional contributions solely from the $\rho$ meson sector when compared to the canonical prescription. The Belinfante energy density $\varepsilon_{\rm Bel}(r)$, mechanical pressure $p_{\rm Bel}^\text{mec}(r)$, and shear force $s_{\rm Bel}(r)$ are expressed as~\cite{Fukushima2026}:

\begin{eqnarray}   
\varepsilon_\text{Bel}(r) &=& \varepsilon_{can}(r)+ 2 f_\pi^2 g^2 \omega^2 + \omega^{\prime 2}
- \frac{3 g\,\omega}{4\pi^2 r^2} F^\prime \sin^2 F \nonumber \\
&=& \frac{1}{2}f_\pi^2\left({F^\prime}^2 + 2\frac{\sin^2 F}{r^2}\right) + f_\pi^2m_\pi^2(1 - \cos F) \nonumber \\
&\quad& + \frac{2}{r^2}f_\pi^2(G + 1 - \cos F)^2 + f_\pi^2g^2\omega^2  \nonumber\\
&\quad&+ \frac{1}{2g^2r^4}\left[2r^2{G^\prime}^2 + G^2(G + 2)^2\right] + \frac{1}{2} {\omega^\prime}^2\,, \label{eq:eBel} 
\end{eqnarray}
\begin{eqnarray}
p_\text{Bel}^\text{mec}(r) &=& p_{\rm can}(r)+\frac{4 f_\pi^2}{3 r^2}\,G\bigl(G+1-\cos F\bigr)\nonumber \\
&\quad& - \frac{2 f_\pi^2}{3 r^2}\,G^2 + \frac{1}{3 g^2 r^2} G^{\prime 2}
+ \frac{1}{3 g^2 r^4} G^2 (G+2) \, \nonumber \\
&=& -\frac{1}{6}f_\pi^2\left({F^\prime}^2 + 2\frac{\sin^2 F}{r^2}\right) - f_\pi^2m_\pi^2(1 - \cos F)\nonumber \\
&\quad&  - \frac{2}{3r^2}f_\pi^2(G + 1 - \cos F)^2 + f_\pi^2g^2\omega^2 \nonumber\\
&\quad& + \frac{1}{6g^2r^4}\left[2r^2{G^\prime}^2 + G^2(G + 2)^2\right] + \frac{1}{6}{\omega^\prime}^2\, , \label{eq:pBel}    
\end{eqnarray}
\begin{eqnarray}
s_{\mathrm{Bel}}(r)&=&s_{\mathrm{can}}(r)-\Bigg[\frac{2 f_\pi^2}{r^2}\, G(G+1-\cos F)\nonumber \\
&\quad& + \frac{1}{g^2 r^4}\Bigg(r^2 G'^2 - 3rGG' + G^2(G+1)(G+2)\Bigg)\Bigg] \nonumber \\
 &=& f_\pi^2\left({F^\prime}^2 - \frac{\sin^2 F}{r^2}\right) - \frac{2}{r^2}f_\pi^2(G + 1 - \cos F)^2 \nonumber \\
&\quad& + \frac{1}{g^2r^4}\left[r^2{G^\prime}^2 - G^2(G + 2)^2\right] - {\omega^\prime}^2\, .\label{eq:shear_Bel}    
\end{eqnarray}

Note that for the two prescriptions of the EMT, we have the following expression for the static symmetric EMT~\cite{Lorce2018,Polyakov2018}:
\begin{equation}
T^{ij}(r)=\delta^{ij} p(r) + \left( \frac{r^i r^j}{r^2}-\frac 1 3 \delta^{ij}\right)s(r) \, .
\end{equation}

The radial and tangential pressures are defined as follows~\cite{Lorce2018}:
\begin{align}
p^{\rm mec}_{r,\,\mathrm{can/Bel}} &= p^{\rm mec}_{\mathrm{can/Bel}} + \frac{2}{3} s_{\mathrm{can/Bel}}\,, \label{eq:pr_mec}\\
p^{\rm mec}_{t,\mathrm{can/Bel}} &= p^{\rm mec}_{r,\,\mathrm{can/Bel}} - s_{\mathrm{can/Bel}}\,. \label{eq:pt_mec}
\end{align}

Furthermore, the conservation of the total energy-momentum tensor (EMT), $\partial_{\mu}T^{\mu\nu}=0$, can be reduced to the following  relation:
\begin{eqnarray}\label{eq:EMT_conservation}
\hat{r}^i \left[\frac{d}{dr}{\left[p^{\rm mec}_{r,\, \rm can/Bel}(r)\right]}+ \frac{2}{r}\,s_{\rm can/Bel}(r)\right]= 0 \,.
\end{eqnarray}
In the absence of singularity at the origin the above Eq.~\eqref{eq:EMT_conservation}, reduces to the hydrostatic equilibrium relation~\cite{Lorce2018}:
\begin{eqnarray}\label{eq:Hydro_equilibrium}
\frac{d}{dr}{\left[p^{\rm mec}_{r,\, \rm can/Bel}(r)\right]}+ \frac{2}{r}\,s_{\rm can/Bel}(r)= 0 \,.
\end{eqnarray}

The existing singularity within the canonical prescription for the EMT prevents Eq.~\eqref{eq:Hydro_equilibrium} from being satisfied near the center of the soliton.

In Ref.~\cite{Lorce2018}, both the isotropic pressure $p^{\rm mec}(r)$ and the radial pressure $p_r^{\rm mec}(r)$ are considered as candidates for describing uniform quark matter properties, and they will be investigated in our analysis. It is also suggested that the radial pressure $p_r$, which represents the normal stress, is the best candidate to contain the bulk isotropic contribution to the pressure in uniform matter. Note that when the shear force vanishes, in uniform matter or in the center of the nucleon, except for the canonical pressure within the $\pi+\omega+\rho$ soliton model, we have $p_r(r)=p(r)$; see Eq.~\eqref{eq:pr_mec}. The tangential pressure $p_t$ is associated with an anisotropic equation of state; see Refs.~\cite{Horvat_2011,Bowers1974,Khunt2025,Mohanty2023}. Since we aim to extract the EoS for dense, uniform and isotropic matter~\cite{Komoltsev2023,Komoltsev2025,Somasundaram:2023a,Somasundaram:2023b}, $p_t$ will not be considered further in our analysis.

We also consider a heuristic mixed approach in which the energy density is obtained from the canonical energy density while the mechanical pressure is obtained from the Belinfante mechanical pressure, as employed in Refs.~\cite{Fukushima2020,Pradhan2025}:
\begin{align}
\varepsilon_\text{mix}(r) &\equiv \varepsilon_\text{can}(r)\, , \label{eq:emix} \\
p_\text{mix}^\text{mec}(r) &\equiv  p_\text{Bel}^\text{mec}(r) \, ,\label{eq:pmix} \\
p_{r,\,\text{mix}}^\text{mec}(r) &\equiv  p_{r,\,\text{Bel}}^\text{mec}(r) \, .\label{eq:prmix} 
\end{align}
There is no theoretical reason to consider this pragmatic approach, except to connect with results from previous papers~\cite{Fukushima2020,Pradhan2025}.

\begin{figure*}[t]
\begin{subfigure}{.32\textwidth}
\centering
\includegraphics[width=\linewidth]{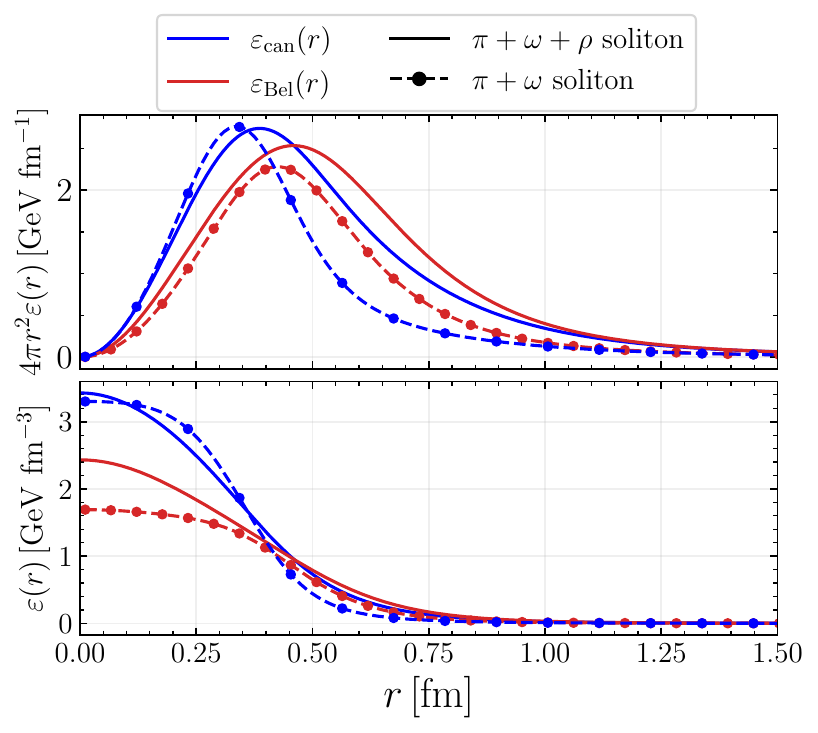}
\caption{}
\label{fig:Energy_pion_omega_rho}
\end{subfigure}
\begin{subfigure}{.32\textwidth}
\centering
\includegraphics[width=\linewidth]{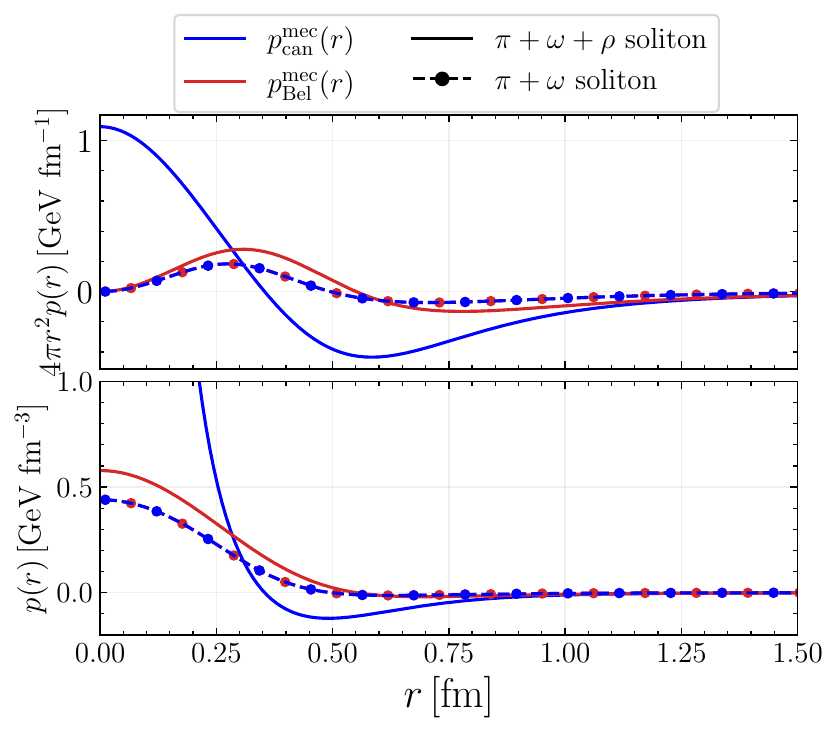}
\caption{}
\label{fig:press_pion_omega_rho}
\end{subfigure}
\begin{subfigure}{.32\textwidth}
\centering
\includegraphics[width=\linewidth]{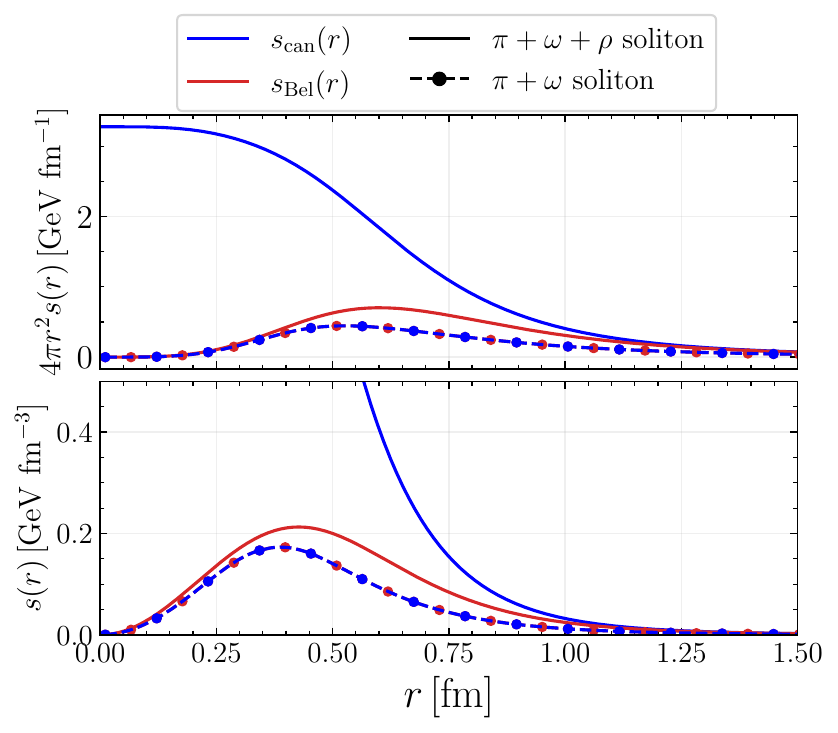}
\caption{}
\label{fig:shear_can_Bel}
\end{subfigure}
\caption{(a) Energy density $\varepsilon_{\rm can}$ and $\varepsilon_{\rm Bel}$ profiles as a function of the radius $r$ (bottom), and energy density distributions for the same quantities (top).
(b) Same as (a) but for $p_{\rm can}^{\rm mec}$ and $p_{\rm Bel}^{\rm mec}$ distributions and profiles. 
(c) Same as (a) but for the shear distributions and profiles.
Results obtained from the canonical EMT are shown in blue and those from the Belinfante EMT in red. For comparison, the profiles of the $\pi+\omega+\rho$ soliton model are shown in solid lines, while the dashed lines with markers correspond to the $\pi+\omega$ soliton model.}
\label{fig:Energy_press_pion_omega_rho}
\end{figure*}

For any prescription defining the energy density (canonical or Belinfante), the soliton mass can be determined in the following way:
\begin{equation}
M_\text{can/Bel} = \int \!\!d^3r \, \varepsilon_\text{can/Bel}(r) \, .
\label{eq:mass}
\end{equation}

The nucleon mass obtained from Eq.~\eqref{eq:mass} is usually larger than the measured one. It was suggested that the reason must probably lie in the fact that the Skyrmion model provides a classical solution for the nucleon, which still needs to be quantized~\cite{Fukushima2020}. A complete treatment requires collective quantization, which generates the nucleon-Delta ($N$-$\Delta$) mass splitting through the moment of inertia and removes spurious rotational contributions. In particular, spin--isospin quantization increases the soliton mass by $3/(8I)$, where $I$ is the moment of inertia. Additionally, one must account for center-of-mass corrections to obtain the physical nucleon mass. 

A parameter $\chi_{\rm can/Bel}$, defined in the following way~\cite{Fukushima2020,Pradhan2025},
\begin{equation}\label{eq:chi}
\chi_{\mathrm{can/Bel}}\equiv \frac{M_N\sim 939\ \mathrm{MeV}}{M_{\mathrm{can/Bel}}}\, .
\end{equation}
has been introduced to incorporate these effects in the equation of state in a phenomenological way. For reasons that will become clear in quark matter, the parameter $\chi_{\rm can/Bel}$ is often named a scaling parameter. Since the soliton masses derived from the canonical and Belinfante EMTs are almost equal\footnote{Numerically, the values we obtain for $\chi_{\mathrm{can}}$ and $\chi_{\mathrm{Bel}}$ differ only at the sixth digit.}, we omit the subscripts and simply write in the following $\chi=\chi_{\mathrm{can}}=\chi_{\mathrm{Bel}}$. 

\subsection{Internal properties of the nucleon}

The internal properties of the nucleon, within the soliton model, are determined from Eqs.~\eqref{eq:ecan}-\eqref{eq:shear_can} for the canonical prescription and Eqs.~\eqref{eq:eBel}-\eqref{eq:shear_Bel} for Belinfante's one. The energy density, the mechanical pressure, and the shear force are shown in Fig.~\ref{fig:Energy_press_pion_omega_rho} for the two prescriptions and for the same field profiles shown in Fig.~\ref{fig:field_profiles}. 

First, we discuss the energy density profiles (bottom) and the energy density distributions (top) for the two prescriptions (canonical and Belinfante) and for the two soliton models ($\pi+\omega$ and $\pi+\omega+\rho$) in Fig.~\ref{fig:Energy_pion_omega_rho}. Despite the difference between the distributions of $\varepsilon_{\rm can} (r)$ and $\varepsilon_{\rm Bel} (r)$ for the two soliton models, the mass of the soliton, see Eq.~\ref{eq:mass}, is found to be identical, $\simeq 1450$ (resp. 1046) MeV, as mentioned before for the $\pi+\omega+\rho$ (resp. $\pi+\omega$) model. For the two soliton models, the energy density is larger (smaller) in the center (external region) of the nucleon for the canonical prescription compared to Belinfante's one. For the same density, see Fig.~\ref{fig:baryon_density_vsr}, the energy density is larger for the canonical prescription compared to Belinfante's one in ultra-dense matter (of the order of 1 to 2~fm$^{-3}$) while it is smaller at relatively lower density (below 1~fm$^{-3}$).

The mechanical pressures are shown in Fig.~\ref{fig:press_pion_omega_rho}. The canonical pressure $p_{\rm can}^{\rm mec}(r)$ exhibits a $r^{-2}$ singular behavior at the center of the nucleon for the $\pi+\omega+\rho$ model, which is absent for the $\pi+\omega$ model. The onset of the singularity at the center is a consequence of the presence of the $\rho$ field, as it will be analyzed in the following. 

The profiles and distributions for the shear force are shown in Fig.~\ref{fig:shear_can_Bel}. For the $\pi+\omega+\rho$ soliton model, the canonical shear force exhibits an $r^{-2}$ singular behavior near the center of the soliton, similar to the pressure singularity shown in Fig.~\ref{fig:press_pion_omega_rho}.
This singularity in both the shear force and pressure within the canonical framework prevents the hydrostatic equilibrium Eq.~\eqref{eq:Hydro_equilibrium} from being satisfied near the origin. In contrast, the hydrostatic equilibrium condition~\eqref{eq:Hydro_equilibrium} is satisfied throughout the soliton volume under the Belinfante prescription. 

\begin{figure}[t]
\centering
\includegraphics[width=\linewidth]{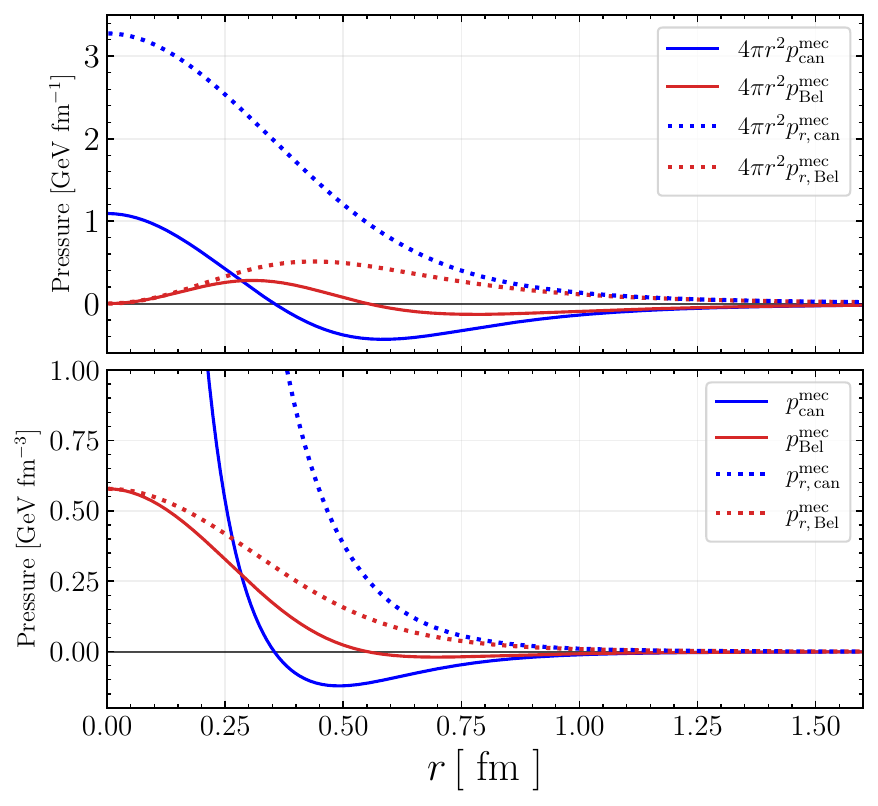}
\caption{Comparison between the radial and mechanical (isotropic) pressure profiles (bottom) and pressure distributions (top) for the canonical and Belinfante prescriptions for the EMT and for the $\pi+\omega+\rho$ soliton model. See text for more details.}
\label{fig:Pmec_Pth_vs_r_compare}
\end{figure}

From the mechanical pressure and the shear force, it is possible to construct the radial and the tangential pressures~\cite{Lorce2018}, see Eqs.~\eqref{eq:pr_mec} and \eqref{eq:pt_mec}. The radial and mechanical (isotropic) pressures are shown in Fig.~\ref{fig:Pmec_Pth_vs_r_compare} for the $\pi+\omega+\rho$ soliton model and for the two prescriptions (canonical and Belinfante). Note that the solid lines in Fig.~\ref{fig:Pmec_Pth_vs_r_compare} are the same as the ones in Fig.~\ref{fig:press_pion_omega_rho}. By construction, the radial pressure has the same singularity as the mechanical one. We also find that the radial pressure exceeds the mechanical pressure throughout the soliton volume. Since the difference between the mechanical and radial pressure is proportional to $s$, it is only for the Belinfante prescription that $p_r^{\rm mec}\sim p^{\rm mec}$ at the center of the nucleon, since $s_\mathrm{Bel}(r\sim 0)\sim 0$. Moreover, the radial pressure $p_r^{\rm mec}(r)$ is positive for all $r$, as required by the mechanical stability~\cite{Polyakov2018,Lorce2018}.

The presence of a singularity near the center of the soliton in the $\pi+\omega+\rho$ model and the canonical prescription may indicate that soliton models with the $\rho$ field should not be considered. The analysis performed in Ref.~\cite{IGARASHI1985}, however, indicates that soliton models stabilized solely by the $\omega$ meson correspond to a saddle-point solution. In other words, the soliton models without the $\rho$ field are not stable. The stability requirement therefore indicates that soliton models must consider the $\rho$ field. Furthermore, it is possible to derive the chiral soliton model from an underlying NJL model by considering a gradient expansion framework~\cite{Zuckert1995,Zuckert1995b}. Such a chiral soliton model naturally generates a $\rho$ field. Finally, the $\rho$ meson plays an important role in describing isospin-asymmetric matter, as encountered in NSs. Hence, in view of the importance of the $\rho$ field, we will investigate in the following only soliton models stabilized by both $\omega$ and $\rho$ mesons, i.e., $\pi+\omega+\rho$ soliton, and disregard in the following the simpler $\pi+\omega$ soliton model. By using the $\pi+\omega+\rho$ soliton model, we, however, face a singularity in the mechanical pressure, which is shown in Fig.~\ref{fig:Energy_press_pion_omega_rho} and commented on previously. The discussion of the origin of the singularity and solutions to circumvent are addressed in the following.

\subsection{Singularity in the pressure and shear force}
\label{sec:singularity}

The energy density originating from the canonical EMT, see Eq.~\eqref{eq:ecan}, can be decomposed into several terms with physical significance: the $\pi$ and $\omega$ masses ($\varepsilon_{\pi M}$ and $\varepsilon_{\omega M}$), the $\pi$ and $\omega$ kinetic energies ($\varepsilon_{\pi K}$ and $\varepsilon_{\omega K}$), the coupling between $\pi$ and $\omega$ ($\varepsilon_{\omega \pi}$), $\pi$-$\rho$ interaction term ($\varepsilon_{\rho \pi}$) and finally the intrinsic $\rho$ Yang-Mills term ($\varepsilon_{\rho\mathrm{YM}}$) containing the explicit gradient term ($\varepsilon_{G^{\prime}}$) and the self-interaction term ($\varepsilon_{GG}$). All these terms can be expressed in the following way:

\begin{widetext}
\begin{align}
\varepsilon_\text{can}(r) &= 
\underbrace{f_\pi^2 m_\pi^2 (1 - \cos F)}_{\varepsilon_{\pi M}}
\underbrace{- f_\pi^2 g^2 \omega^2}_{\varepsilon_{\omega M}} 
+\underbrace{\frac{1}{2} f_\pi^2 \left(F'^2 + 2 \frac{\sin^2 F}{r^2}\right)}_{\varepsilon_{\pi K}}
\underbrace{- \frac{1}{2} \omega'^2}_{\varepsilon_{\omega K}}
 +\underbrace{\frac{3 g \omega}{4 \pi^2 r^2} F' \sin^2 F}_{\varepsilon_{\omega \pi}}
\nonumber \\
&+ \underbrace{\frac{2 f_\pi^2}{r^2} (G+1-\cos F)^2}_{\varepsilon_{\rho \pi}} + \underbrace{
\underbrace{\frac{G'^2}{g^2 r^2}}_{\varepsilon_{G^{\prime} }} + \underbrace{\frac{G^2 (G+2)^2}{2 g^2 r^4}}_{\varepsilon_{GG}}}_{\varepsilon_{\rho {\rm YM}}}.
 \end{align}
A similar decomposition can be done for the mechanical pressure originating from the canonical EMT:
\begin{align}
p_\text{can}^\text{mec}(r) &= 
\underbrace{- f_\pi^2 m_\pi^2 (1 - \cos F)}_{p_{\pi M}}
+\underbrace{ f_\pi^2 g^2 \omega^2}_{p_{\omega M}}
\underbrace{- \frac{1}{6} f_\pi^2 \left(F'^2 + 2 \frac{\sin^2 F}{r^2} \right)}_{p_{\pi K}}
+\underbrace{\frac{1}{6} \omega'^2}_{p_{\omega K}} 
+ \underbrace{0}_{p_{\omega \pi}} 
\nonumber \\
&
\underbrace{- \frac{2 f_\pi^2}{3 r^2} (3G+1 - \cos F)(G + 1 - \cos F)}_{p_{\rho\pi}}
\underbrace{\underbrace{-\frac{G'^2}{3 g^2 r^2} }_{p_{G^{\prime}}} 
\underbrace{-\frac{G^2(G+2)}{6 g^2 r^4} \left[3 G+2\right]}_{p_{GG}}}_{p_{\rho {\rm YM}}}.
\end{align}
\end{widetext}
\begin{figure*}[t]
\centering
\begin{subfigure}{.49\textwidth}
  \centering
  \includegraphics[width=0.96\linewidth]{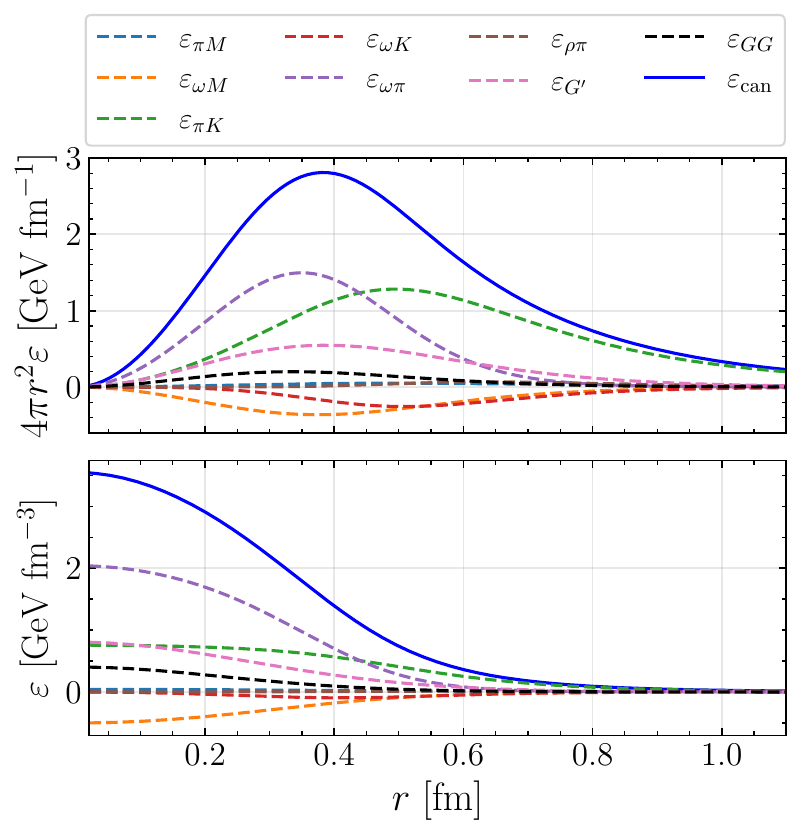}
  \caption{}
  \label{fig:Ecan_component}
\end{subfigure}
\begin{subfigure}{.49\textwidth}
  \centering
\includegraphics[width=\linewidth]{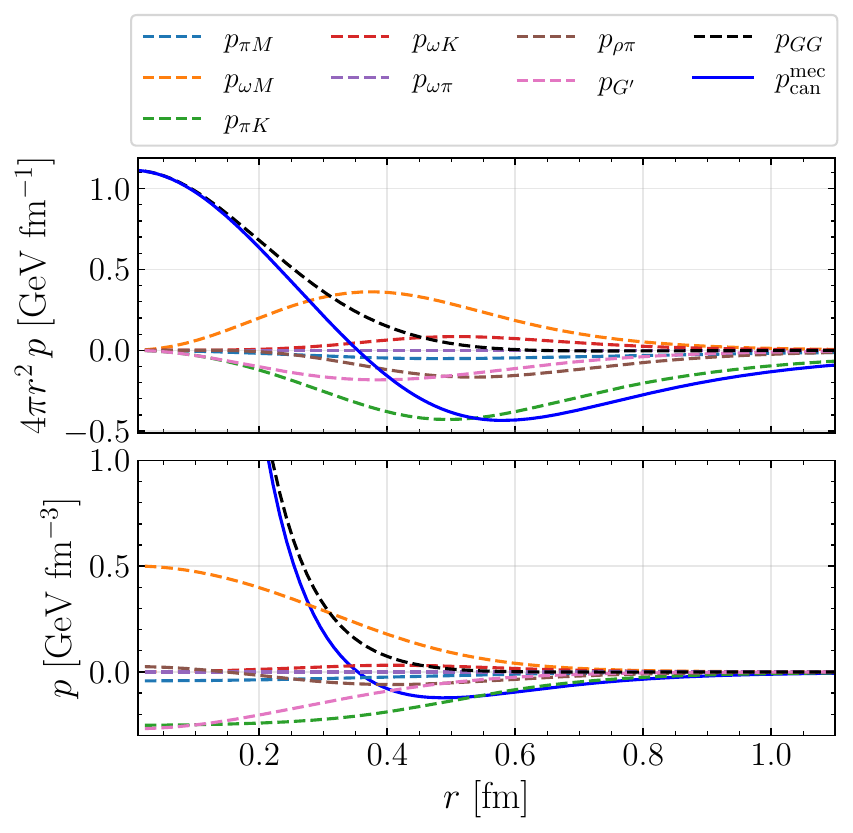}
  \caption{} 
  \label{fig:pcan_component}
\end{subfigure}%
\caption{(a) Energy distributions  (top) and energy density profiles (bottom) corresponding to different components of $\varepsilon_{\rm can}(r)$ (a) as a function of $r$ for the $\pi+\omega+\rho$ soliton model. (b) Same as (a) but for $p_{\rm can}^{\rm mec}(r)$~. For better comparison and visibility, the radial axes are limited to a certain maximum value.}
\label{fig:Ecan_pcan_component}
\end{figure*}

These various terms are displayed in Fig.~\ref{fig:Ecan_component} and \ref{fig:pcan_component}. These contributions can be directly compared with analogous decompositions appearing in different models of the nucleon (for instance, the one discussed in Refs.~\cite{Chanfray2026a,Chanfray2026b}). The largest contributions to the energy density are generated by the $\pi$ terms, more specifically by $\varepsilon_{\pi K}$ and $\varepsilon_{\omega \pi}$. The largest contributions to the pressure are generated by the $\pi$ and $\rho$ terms, more specifically by $p_{\pi K}$ and $p_{GG}$. The $\pi$ contribution, more specifically the $\varepsilon_{\pi K}$ and $p_{\pi K}$ terms, governs the asymptotic behavior of the energy density and pressure, respectively. The singularity observed in $p^{\rm mec}_{\rm can}$ originates from the $\rho$-meson sector, especially from the YM self-interaction term ($p_{GG}$) contribution to the YM term $p_{\rho \rm {YM}}$. It is therefore clear that the origin of the singularity is the $\rho$ field.

As illustrated in Fig.~\ref{fig:pcan_component}, the singular behavior of the canonical pressure near the soliton center is driven by the sole contribution of the $p_{GG}$ term. This feature can be understood directly by examining the small-$r$ behavior of the field $G(r)$ provided by Eq.~\eqref{eq:G_small_r}. The numerators of $p_{GG}$, scale as $r^{2}$, whereas the denominator as $r^{4}$, and when combined, the explicit $(G+2)/r^{4}$ term produces a $1/r^{2}$ divergence in the canonical pressure $p_{\rm can}^{\rm mec}$ at the soliton center. In contrast, the $\rho$-meson contribution to the Belinfante EMT, $p_{\rm Bel}^{\rm mec}$, always appears in the combination $(G+2)/r^{2}$, see Eq.~\eqref{eq:pBel}. This weaker $r^{-2}$ pre-factor is canceled by the $r^{2}$ vanishing of $G+2$ near $r\to 0$, yielding regular behavior of $p_{\rm Bel}^{\rm mec}$ as $r\to0$, clearly visible in Fig.~\ref{fig:Energy_press_pion_omega_rho}.

The singular behavior of $s_{\rm can}(r)$ can be understood in a similar manner. Its origin is also due to the presence of terms proportional to $(G+2)/r^{4}$ in the canonical shear force.

\subsection{Analogy with the liquid-drop model}
\label{sec:ldm}

The liquid-drop model for nuclei~\cite{Gamow:1928} is based on the leptodermous expansion~\cite{Myers:1973}, introducing a bulk contribution to the energy in finite-size nuclei, a surface contribution, etc. It was suggested in Ref.~\cite{Polyakov2018} to draw an analogy between the energy of the nucleon and the liquid-drop model to identify bulk and surface contributions: An average surface tension $\gamma$ and the corresponding surface energy $E_\mathrm{surf}$ of the soliton were defined as,
\begin{eqnarray}
\gamma_{\rm can/Bel} &\equiv& \frac{1}{4\pi}\int \!\!d^3r \, \frac{s_{\rm can/Bel}(r)}{r^2}\,, \label{eq:gamma}\\
E_{\rm surf,\,can/Bel} &\equiv& \int \!\!d^3r \, s_{\rm can/Bel}(r)\,. \label{eq:surface_energy}
\end{eqnarray}
This analogy is mostly justified by the fact that the shear force is localized around the surface. It is indeed observed in Fig.~\ref{fig:shear_can_Bel} for the $\pi+\omega$ soliton model, and for the Belinfante prescription within the $\pi+\omega+\rho$ soliton model. However, it is not the case for the canonical prescription within the $\pi+\omega+\rho$ soliton model, which is related to the onset of a singularity at the center of the nucleon. This therefore limits the suggested analogy.

Using the mechanical properties shown in Fig.~\ref{fig:shear_can_Bel} and Eq.~\eqref{eq:surface_energy}, the surface energies are estimated to be $E_{\rm surf,\,can} \sim 2250~\mathrm{MeV}$ and $E_{\rm surf,\,Bel} \sim 485~\mathrm{MeV}$ for the $\pi+\omega+\rho$ model, respectively. The corresponding Belinfante surface tension is found to be $\gamma_{\rm Bel} \sim 114$ MeV~fm$^{-2}$~\footnote{The canonical shear force $s_{\rm can}(r)$ leads to a divergent integral for $\gamma_{\rm can}$ due to its singular behavior near the core of the soliton.}. 

When the shear force is localized around the nucleon's surface, that is, for all cases except for the canonical prescription in the $\pi+\omega+\rho$ soliton model, the following approximations are assumed~\cite{Polyakov2018}:
\begin{eqnarray}
s_{\rm can/Bel}(r) &\approx& \gamma_{\rm can/Bel} \, \delta(r - R_{\rm drop}) \, , \label{eq:sapprox} \\
E_{\rm surf,\,can/Bel} &\approx& 4\pi \gamma_{\rm can/Bel} R_{\rm drop}^2 \, , \label{eq:esurfapprox}
\end{eqnarray}
where $R_{\rm drop}$ is defined as being the droplet radius: $2\gamma / R_{\rm drop} = p^{\rm mec}(r=0)$~\cite{Polyakov2018}.

The relations~\eqref{eq:sapprox} and \eqref{eq:esurfapprox} are more and more accurate as the shear force $s(r)$ is more and more localized at the nucleon's surface. In other words, these relations assume a sharp density variation confined to a thin surface region. However, Fig.~\ref{fig:shear_can_Bel} shows that $s$ is non-zero over an extended region of the soliton volume in all cases, except the canonical prescription in the $\pi+\omega+\rho$ soliton model, rather than being localized within a narrow surface layer. This raises the question of the liquid-drop model analogy of the soliton as stated in Ref.~\cite{Polyakov2018}.

Still assuming the analogy suggested in Ref.~\cite{Polyakov2018}, the bulk energy density is obtained as
\begin{equation}
\varepsilon^{\rm Bulk}_{\rm can/Bel}\equiv\varepsilon_{\rm can/Bel}-\varepsilon^{\rm Surf}_{\rm can/Bel}
\, ,
\label{eq:ebulk}
\end{equation}
with $\varepsilon^{\rm Surf}_{\rm can/Bel}\equiv s_{\rm can/Bel}$. For the cases where this analogy may be accepted, i.e., all cases except the canonical prescription in the $\pi+\omega+\rho$ soliton model, the quantity $\varepsilon^{\rm Bulk}_{\rm can/Bel}$ obtained from \eqref{eq:ebulk} is positive over the entire volume of the nucleon. As a consequence, the total bulk energy is positive. In the liquid-drop model, the bulk contribution is often negative, the surface is positive, and the stability of the liquid drop is mainly obtained from the competition of these two contributions to the total energy. The positivity of the bulk energy that we found is therefore another reason to question the analogy suggested in Ref.~\cite{Polyakov2018}.

Finally, the analogy suggested in Ref.~\cite{Polyakov2018} does not define the bulk contribution to the pressure, which could be employed in uniform matter. In the present case, we therefore apply the analogy suggested in Ref.~\cite{Polyakov2018} only to the energy. We, however, note that a bulk contribution to the pressure, where the surface contribution is removed, would be necessary to map the nucleon's properties to the EoS in supra-dense matter.

\section{Quark Matter Equation of State}
\label{sec:qm}

We now address the question of the construction of the quark matter (QM) EoS from nucleon models, assuming a direct mapping between the nucleon properties from the soliton model and the QM EoS. 

In uniform matter, the thermodynamic pressure is defined as
\begin{equation}
p^\mathrm{th} \equiv n^2 \frac{\partial \varepsilon/n}{\partial n} \, ,
\label{eq:pth}
\end{equation}
where $n$ and $\varepsilon$ are the particle and energy densities in uniform matter. Considering a direct mapping between nucleon and uniform matter, they are obtained from the density and the energy density profile in the soliton model, for instance, $n\equiv n_B(r)$ shown in Fig.~\ref{fig:baryon_density_vsr}. In uniform matter, the thermodynamic~\eqref{eq:pth} and the mechanical canonical~\eqref{eq:pcan} and Belinifante \eqref{eq:pBel} pressures must coincide. In the following, we show that it is not the case.

\subsection{From nucleon models to quark matter}

\begin{figure}[t]
\centering
\includegraphics[width=\linewidth]{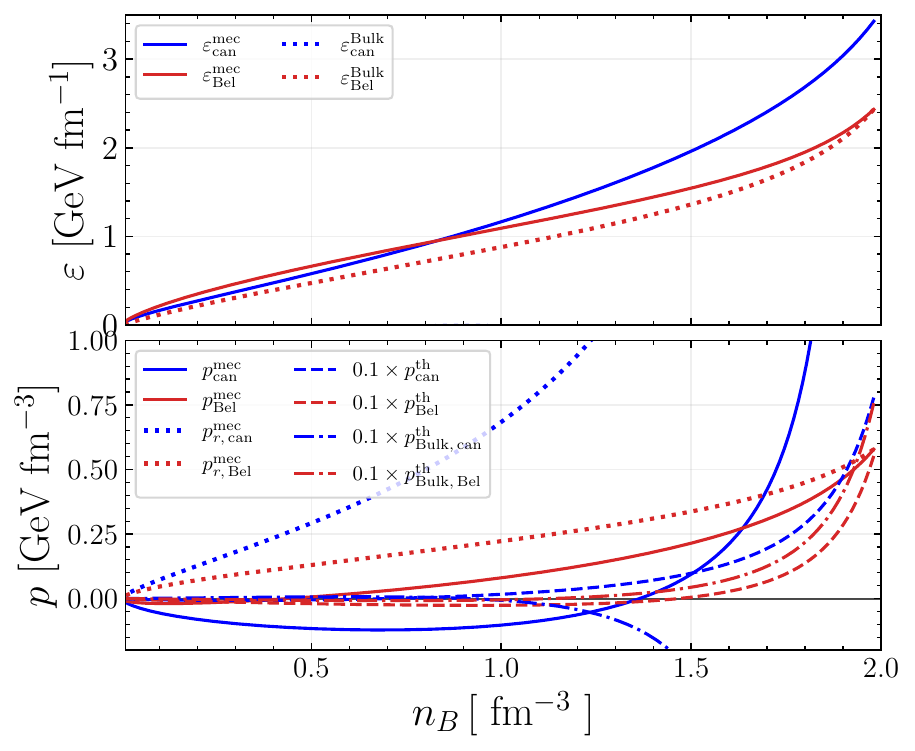}
\caption{Energy density $\epsilon$ (top) and pressure $p$ (bottom) as a function of the baryon density $n_B$ for the different models considered here: canonical, Belinfante, mechanical, and bulk. The bulk energy density from the canonical EMT is $\varepsilon_{\rm can}^{\rm Bulk}\leq 0$ for all densities, therefore, it does not appear in the figure. For the pressure, we additionally show the thermodynamic pressure as defined in Eq.~\eqref{eq:pth} for the canonical and Belinfante EMT.}
\label{fig:Pmec_Pth_vs_nb_compare}
\end{figure}

\begin{figure*}[t]
\centering

\begin{minipage}{0.48\textwidth}
    \centering
    \includegraphics[width=\linewidth]{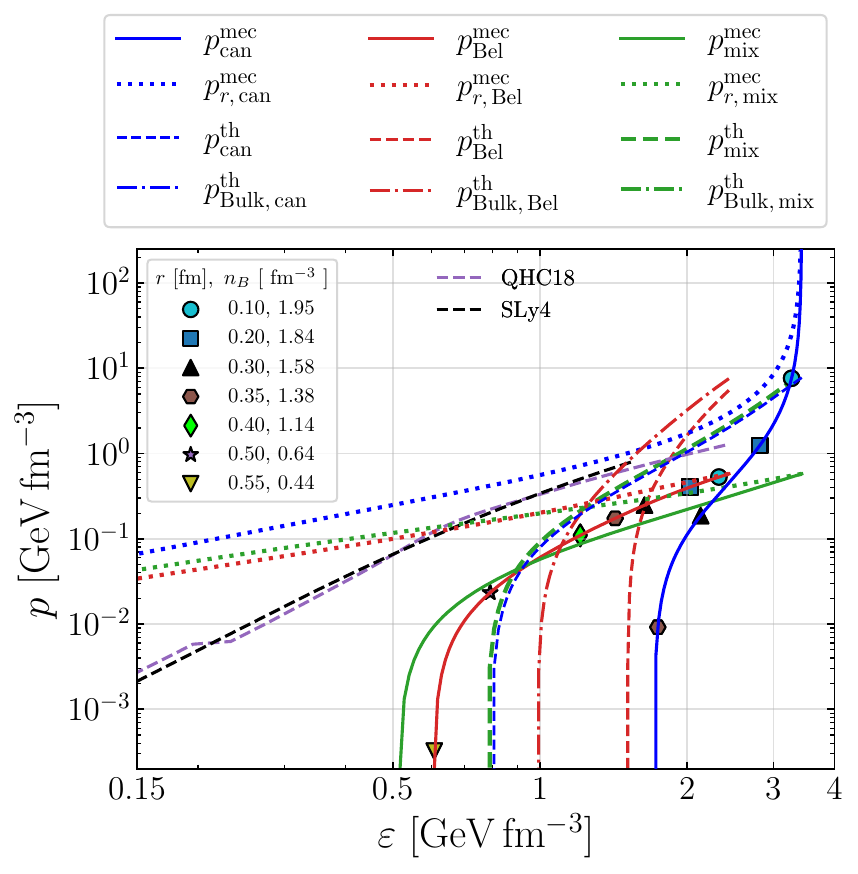}
    \caption{EoSs of dense matter derived from the mechanical properties of the soliton. For comparison, the SLy4~\cite{Gulminelli_2015_SLy4,Chabanat:1998} and QHC18~\cite{Baym_QHC2018} EoSs are also shown. See text for complete details.}
    \label{fig:EoS_compare_mec_thermo}
\end{minipage}
\hfill
\begin{minipage}{0.49\textwidth}
    \centering
    \includegraphics[width=\linewidth]{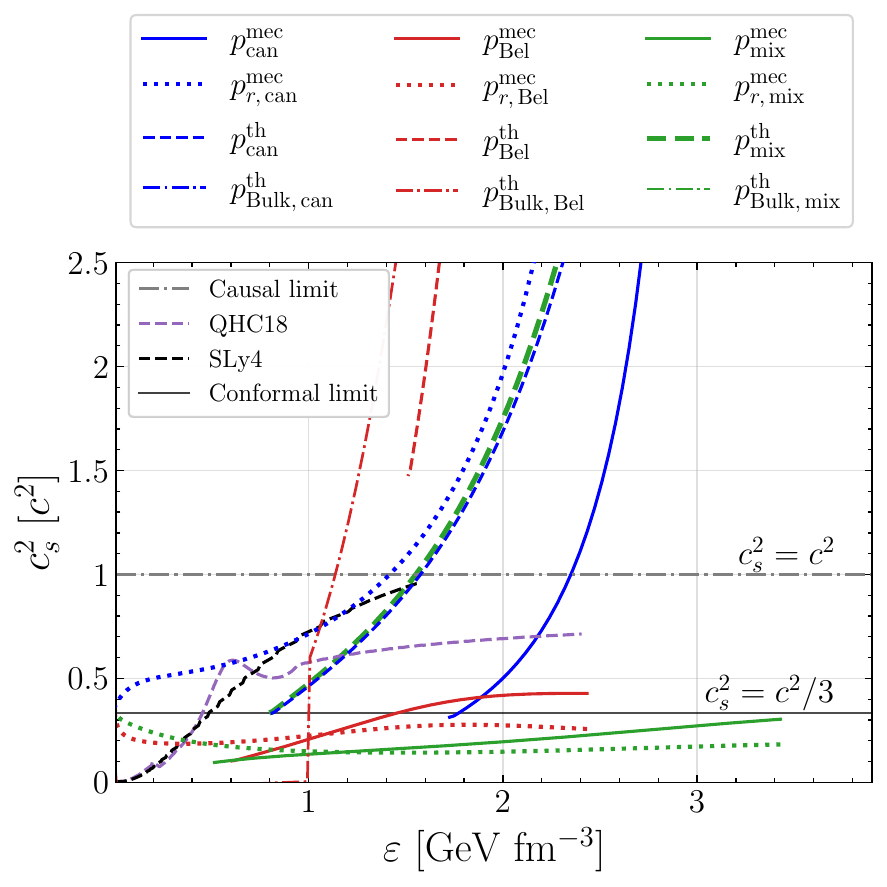}
    \caption{Sound speed $c_s^2$ as a function of the energy density corresponding to the EoSs displayed in Fig.~\ref{fig:EoS_compare_mec_thermo}.}
    \label{fig:speed_of_sound_vs_eb_compare_mec_thermo}
\end{minipage}

\end{figure*}
The energy density as well as the mechanical and thermodynamical pressures for different prescriptions are shown in Fig.~\ref{fig:Pmec_Pth_vs_nb_compare} as a function of the baryon density $n_B$. This figure illustrates how nucleon properties are employed to construct the QM EoS: using the relation between the baryon density $n_B$  and $r$ (see Fig.~\ref{fig:baryon_density_vsr}), the explicit $r$ dependence in the energy density and pressure profiles is used to obtain a functional relation $\varepsilon(n_B)$ and $p(n_B)$. The mapping between soliton properties and the EoS has been suggested in Ref.~\cite{Fukushima2020}, and employed in Refs.~\cite{Pradhan2025,Fukushima2026}. Let us mention that this mapping requires that, locally in the soliton, the mechanical properties represent the same properties as those of ultra-dense matter. In other words, the role of the surface correction in the soliton model is disregarded in this mapping. Moreover, by construction of the mapping, only the densities for which the pressure is positive contribute to the stable EoS. The densities for which the pressure is negative localize a region where the soliton is unstable in ultra-dense matter. The density at which the pressure changes its sign is defined as the limiting density in supra-dense matter. This is the lower density at which the EoS is defined. 

As suggested in Ref.~\cite{Fukushima2026}, the results should be independent of the prescription for the EMT. We add that it should also be independent of the definition employed for the pressure: mechanical versus thermodynamic. In Fig.~\ref{fig:Pmec_Pth_vs_nb_compare}, a significant difference between $p^{\rm mec}_{\rm can/Bel}$ and $p^{\rm th}_{\rm can/Bel}$ persists even in the high-density region corresponding to the center of the soliton, where surface contributions are expected to be smaller than near the surface. This indicates that, most probably, surface terms contribute even in the center of nucleons. This is not surprising since nucleons are formed by a very small number of quarks. The nucleon density does not indicate a clear separation between bulk and surface regions, see Fig.~\ref{fig:baryon_density_vsr}. In nuclei, for instance, only intermediate-mass and heavy nuclei have a central area that resembles uniform matter (the bulk of nuclei), while in light nuclei, there is no region where the density is flat, and where the local density approximation allows one to map nuclear properties and the properties of uniform matter.

Finally, we notice that in Fig.~\ref{fig:Pmec_Pth_vs_nb_compare} there is a qualitative difference in the densities at which the mechanical and thermodynamic pressures change their sign. 

We now analyze the impact of the EMT prescriptions (canonical, Belinfante, and mix) on the EoS (pressure versus energy density) as well as the difference between mechanical and thermodynamic definitions of the pressure. 

EoSs can be constructed in different ways, as shown in Fig.~\ref{fig:EoS_compare_mec_thermo} and in the following summary:
\begin{itemize}
\item $p^{\rm mec}_{\rm can}(\varepsilon_{\rm can})$: $p(\epsilon)$ EoS derived entirely from the canonical EMT $T^{\mu\nu}_{\rm can}$. This is shown as the solid blue line in  Fig.~\ref{fig:EoS_compare_mec_thermo}.
\item $p^{\rm mec}_{\rm Bel}(\varepsilon_{\rm Bel})$: $p(\epsilon)$ EoS derived entirely from the Belinfante EMT $T^{\mu\nu}_{\rm Bel}$. This is shown as the solid red line in  Fig.~\ref{fig:EoS_compare_mec_thermo}.
\item $p^{\rm mec}_{\rm mix}(\varepsilon_{\rm mix})$: the mixed $p(\epsilon)$ EoS constructed using the Belinfante pressure and the canonical energy density, defined in Eqs.~\eqref{eq:emix}-\eqref{eq:pmix}. This is shown as the solid green line in  Fig.~\ref{fig:EoS_compare_mec_thermo}.
\item $p^{\rm mec}_{r,\,\mathrm{can/Bel/mix}}(\varepsilon_{\mathrm{can/Bel/mix}})$: Same as before for $p_r(\epsilon)$ and represented with dotted lines of blue, red, and green colors in  Fig.~\ref{fig:EoS_compare_mec_thermo}.
\item In addition the thermodynamic pressures, $p^{\rm th}_{\rm can}$ and $p^{\rm th}_{\rm Bel}$, are obtained from the corresponding energy densities $\varepsilon_{\rm can}$ and $\varepsilon_{\rm Bel}$, respectively. The resulting EoSs $p^{\rm th}_{\rm can}(\varepsilon_{\rm can})$ and $p^{\rm th}_{\rm Bel}(\varepsilon_{\rm Bel})$ are shown by dashed lines in  Fig.~\ref{fig:EoS_compare_mec_thermo}. 
\item The bulk thermodynamic pressures $p^{\rm th}_{\rm Bulk,\, can}$ and $p^{\rm th}_{\rm Bulk,\, Bel}$, obtained from Eq.~\eqref{eq:pth} using the bulk energy densities defined in Eq.~\eqref{eq:ebulk}, are shown as functions of the corresponding energy densities, i.e., $p^{\rm th}_{\rm Bulk,\, can/Bel}(\varepsilon^{\rm Bulk}_{\rm can/Bel})$, by the dash-dotted lines in Fig.~\ref{fig:EoS_compare_mec_thermo}. Since the bulk energy densities in the canonical and mixed frameworks, $\varepsilon^{\rm Bulk}_{\rm can/mix}\leq 0$ throughout the spatial region, they do not appear in Fig.~\ref{fig:EoS_compare_mec_thermo}.
\end{itemize}

The mechanical Belinfante and the mixed pressure (solid red and solid green) are quite close, since the only difference between these two predictions is the energy density $\varepsilon$ considered in these approaches. The thermodynamic canonical and mixed pressures (dashed blue and dashed green) are identical, since the baryon density $n_B$ is the same and the thermodynamic pressure is obtained from the derivative of the same energy density $\varepsilon_\text{can}$, see Eqs.~\eqref{eq:pth} and \eqref{eq:emix}, which are identical for these two predictions. Note that the associated curves have been slightly shifted for clarity.

Since the EoSs are constructed only from regions where $p>0$, the different EoSs considered here exhibit distinct lower bounds in the $p$-$\varepsilon$ plane. As shown in  Fig.~\ref{fig:Pmec_Pth_vs_r_compare}, the canonical mechanical pressure $p^{\rm mec}_{\rm can}$ becomes $\leq 0$ for $r \gtrsim 0.35\,\mathrm{fm}$, corresponding to $n_B \lesssim 1.4\,\mathrm{fm}^{-3}$, as seen in  Fig.~\ref{fig:Pmec_Pth_vs_nb_compare}. This restricts the lower domain of the EoS constructed from $p^{\rm mec}_{\rm can}$, as indicated in  Fig.~\ref{fig:EoS_compare_mec_thermo}. In contrast, the Belinfante mechanical pressure $p^{\rm mec}_{\rm Bel}$ remains positive up to a larger radius, $r \simeq 0.55\,\mathrm{fm}$, corresponding to lower densities $n_B \sim 0.4\,\mathrm{fm}^{-3}$. For the thermodynamical pressures, a different behavior is observed: $p^{\rm th}_{\rm can}$ vanishes at $r \simeq 0.5\,\mathrm{fm}$, whereas $p^{\rm th}_{\rm Bel}$ reaches zero already at $r \simeq 0.3\,\mathrm{fm}$, as illustrated in  Fig.~\ref{fig:Pmec_Pth_vs_r_compare}.

It is interesting to mention that the radial pressures derived from the EMTs remain positive throughout the density (energy density) regime. Hence, as a result, the EoSs shown as dotted lines in Fig.~\ref{fig:EoS_compare_mec_thermo}, the limiting value corresponding to $\varepsilon\sim 0$.

In Fig.~\ref{fig:speed_of_sound_vs_eb_compare_mec_thermo}, we show the squared speed of sound, $c_s^2=dp/d\varepsilon$, as a function of the energy density $\varepsilon$, for the different EoSs presented in  Fig.~\ref{fig:EoS_compare_mec_thermo}. It is interesting to observe that some EoS become acausal at low energy density. It is the case for $p^\text{th}_\text{Bel}(\varepsilon_\text{Bel})$, which is acausal for all explored densities, as well as for $p^\text{mec}_\text{can}(\varepsilon_\text{can})$ and $p^\text{th}_\text{can}(\varepsilon_\text{can})$, which are causal at low density but become acausal when the energy density exceeds 2.4 and 1.6~GeV fm$^{-3}$ respectively.
It can be remarked that pQCD is predicted to start for $\varepsilon=2.4$~GeV.fm$^{-3}$~\cite{Gorda2023}. We are therefore not far from this limit where the square of the sound speed is expected to be close to the conformal limit ($c_s^2=c^2/3$).

From Fig.~\ref{fig:EoS_compare_mec_thermo} one could conclude that the Belinfante thermodynamic and bulk EoS are rather close, justifying the Belinfante choice for the EMT. However, these two EoSs predict acausal behaviors as shown in Fig.~\ref{fig:speed_of_sound_vs_eb_compare_mec_thermo}.

\subsection{Scaling of the equation of state}
\label{subsec:scaling}
\begin{figure*}[t]
\centering

\begin{minipage}{0.48\textwidth}
    \centering
    \includegraphics[width=\linewidth]{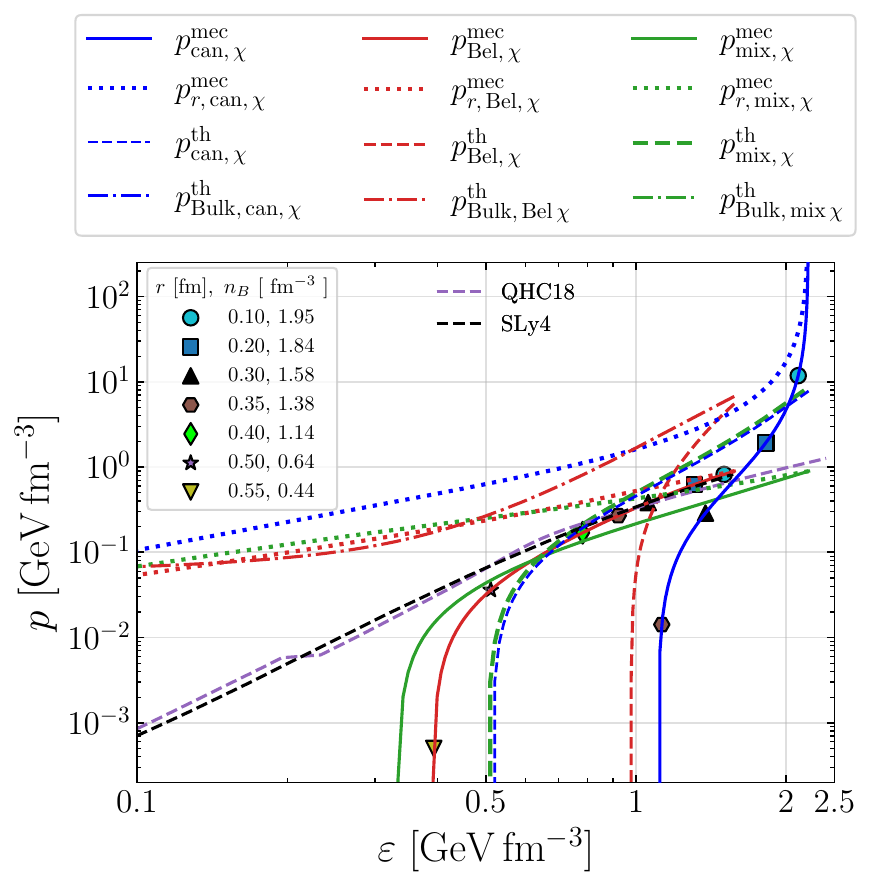}
    \caption{Same as Fig.~\ref{fig:EoS_compare_mec_thermo}, but for the modified EoSs including the scaling factor $\chi$.}
    \label{fig:EoS_compare_mec_thermo_with_scaled}
\end{minipage}
\hfill
\begin{minipage}{0.49\textwidth}
    \centering
    \includegraphics[width=\linewidth]{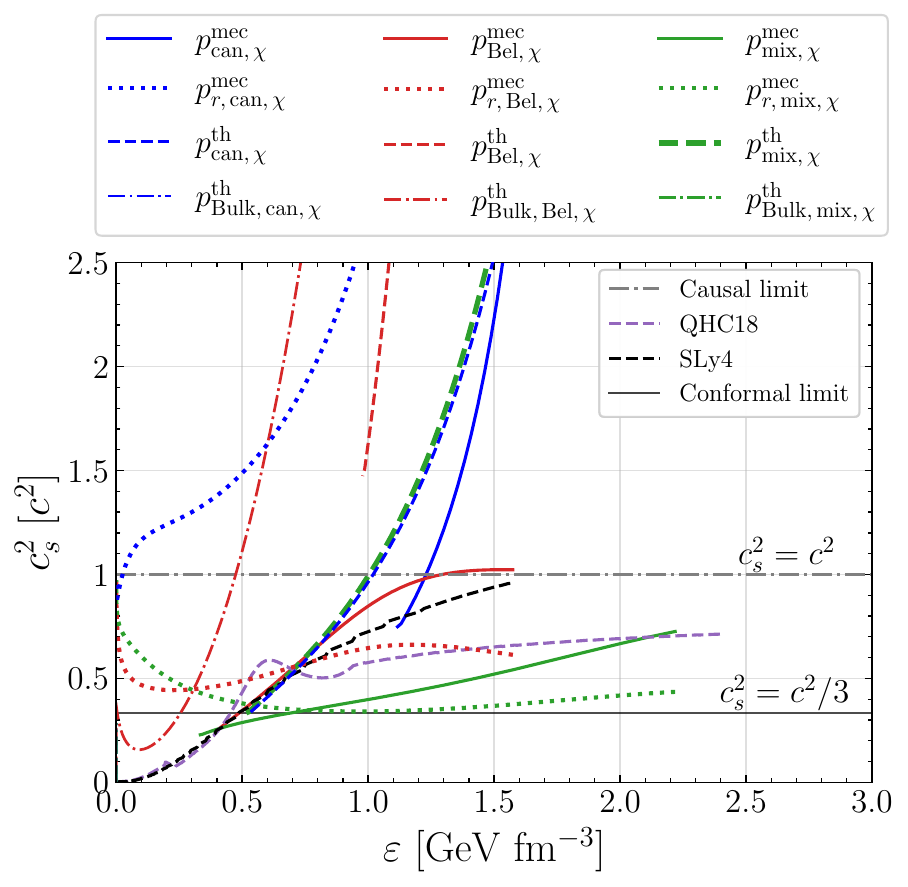}
    \caption{Sound speed $c_s^2$ as a function of the energy density corresponding to the EoSs displayed in Fig.~\ref{fig:EoS_compare_mec_thermo_with_scaled}.}
    \label{fig:speed_of_sound_vs_eb_compare_mec_thermo_scaled}
\end{minipage}
\end{figure*}
It was previously discussed that Skyrmion provides a classical solution that needs to be quantized. Instead of performing explicit quantization, and since its effect is contained in the scaling parameter $\chi$, see Eq.~\eqref{eq:chi}, the energy density is phenomenologically scaled as~\cite{Fukushima2020,Pradhan2025,Fukushima:2025}:
\begin{equation}
\varepsilon_{\mathrm{can/Bel},\,\chi} \equiv \chi \, \varepsilon_{\mathrm{can/Bel}}\, .
\end{equation}
Furthermore, the bulk energy density is defined as,
\begin{equation}
\label{eq:ebulk_chi}
\varepsilon^{\rm Bulk}_{\rm can/Bel,\, \chi}\equiv \varepsilon_{\rm can/Bel,\, \chi}-s_{\rm can/Bel,\,\chi}
\end{equation}

The mechanical pressures and shear forces depend inversely on the soliton mass~\cite{Polyakov2018}, providing the following relations
\begin{align}
p^{\rm mec}_{\mathrm{can/Bel},\,\chi} &\equiv \frac{1}{\chi}\, p^{\rm mec}_{\mathrm{can/Bel}}\, , \label{eq:pmec_chi} \\
p^{\rm mec}_{r,\, \mathrm{can/Bel},\,\chi} &\equiv \frac{1}{\chi}\, p^{\rm mec}_{r, \,\mathrm{can/Bel}}\, , \label{eq:pr_mec_chi}\\
p^{\rm mec}_{t,\, \mathrm{can/Bel},\,\chi} &\equiv \frac{1}{\chi}\, p^{\rm mec}_{t, \,\mathrm{can/Bel}}\, , \label{eq:pt_mec_chi} \\
s_{\mathrm{can/Bel},\,\chi} &\equiv \frac{1}{\chi}\, s^{\rm mec}_{\mathrm{can/Bel}}\, , \label{eq:shear_chi}
\end{align}
and the thermodynamic pressure scales as:
\begin{equation}\label{eq:pth_chi}
p^{\rm th}_{\mathrm{can/Bel},\,\chi} = \chi \, p^{\rm th}_{\mathrm{can/Bel}}\, .
\end{equation}
Note that the scaling factor $\chi$ acts differently for the mechanical pressure, see Eq.~\eqref{eq:pmec_chi}, and the thermodynamic one, see Eq.~\eqref{eq:pth_chi}.

The EoSs with scaling $\chi$, obtained from the mechanical properties of the soliton, i.e., $p^{\rm mec}_{\mathrm{can/Bel},\,\chi}$ as a function of $\varepsilon_{\mathrm{can/Bel},\,\chi}$, are shown as solid lines in Fig.~\ref{fig:EoS_compare_mec_thermo_with_scaled}, where we adopt the same convention as discussed for Fig.~\ref{fig:EoS_compare_mec_thermo}. The effect of the scaling factor $\chi$ can be seen by comparing these results with those in Fig.~\ref{fig:EoS_compare_mec_thermo}, along with the reference SLy4 and QHC18 EoSs. Since $\chi<1$, the rescaled mechanical pressure increases, while the corresponding energy density shifts to lower values. As a result, the EoSs become stiffer compared to those obtained without scaling. The EoSs, $p^\text{th}_{\mathrm{can/Bel},\,\chi}$ as functions of $\varepsilon_{\mathrm{can/Bel},\,\chi}$, are shown as dashed lines in Fig.~\ref{fig:EoS_compare_mec_thermo_with_scaled}. For clarity of visualization, the curve $p^{\rm th}_{\text{mix},\,\chi}(\varepsilon_{\text{mix},\,\chi})$ is slightly shifted in Fig.~\ref{fig:EoS_compare_mec_thermo_with_scaled}, by a similar fashion as done in Fig.~\ref{fig:EoS_compare_mec_thermo}. The effect of the scaling factor $\chi$ is to reduce both the energy density and the thermodynamic pressure, shifting the curves toward lower values along both axes. Although the inclusion of the phenomenological scaling factor $\chi$ increases the mechanical pressure while reducing the thermodynamic pressure, one can still notice clear differences among the different pressures. For the same reason they are not shown in Fig.~\ref{fig:EoS_compare_mec_thermo}, the bulk EoS for the canonical and mixed approaches do not appear in Fig.~\ref{fig:EoS_compare_mec_thermo_with_scaled}. Only the bulk Belinfante EoS appears in Fig.~\ref{fig:EoS_compare_mec_thermo_with_scaled}. It is, however, quite different from the thermodynamic EoS.

In Fig.~\ref{fig:speed_of_sound_vs_eb_compare_mec_thermo_scaled}, we show the squared speed of sound, $c_s^2=dp/d\varepsilon$, as a function of the energy density for the EoSs presented in Fig.~\ref{fig:EoS_compare_mec_thermo_with_scaled}. The increased stiffness of the EoSs derived from mechanical pressure is also reflected in the behavior of the speed of sound. Using Eqs.~\eqref{eq:chi} and~\eqref{eq:pmec_chi}, the speed of sound obtained from the mechanical pressures, i.e, $c_s^2 = d p^{\rm mec}_{i,\,\chi}/d \varepsilon_{i,\,\chi},\ i=\{\mathrm{can,\ Bel,\ mix}\}$~, are enhanced by a factor of $1/\chi^2$ compared to the unscaled case which are shown as in solid lines in Fig.~\ref{fig:speed_of_sound_vs_eb_compare_mec_thermo}. However, with the scaled energy density defined in Eq.~\eqref{eq:chi} and the modified thermodynamic pressure given in Eq.~\eqref{eq:pth_chi}, the corresponding squared speed of sound, $c_s^2 = d p^{\rm th}_{i,\,\chi}/d \varepsilon_{i,\,\chi}, \ i=\{\mathrm{can,\ Bel,\ mix}\}$ displayed with dashed lines in Fig.~\ref{fig:speed_of_sound_vs_eb_compare_mec_thermo_scaled}, remains unchanged compared to  those shown as dashed lines in Fig.~\ref{fig:speed_of_sound_vs_eb_compare_mec_thermo}.

Interestingly, the EoS, $p^\text{mec}_{\text{Bel},\,\chi}(\varepsilon_{\text{Bel},\,\chi})$ becomes acausal in a certain density range ($\varepsilon \sim 1.5~\mathrm{GeV\,fm^{-3}}$), whereas the unscaled EoS $p^{\rm mec}_{\rm Bel}(\varepsilon_{\rm Bel})$, remains causal over the entire energy density range.

\section{Conclusions}
\label{sec:conclusions}

In this work, we have considered the $\pi+\omega$ and $\pi+\omega+\rho$ soliton models, and we have confirmed the result that there is a singularity in the mechanical pressure for the $\pi+\omega+\rho$ model~\cite{Fukushima2026}. The singularity originates from the so-called $GG$ contribution to the YM term. The soliton model including $\rho$ is the same as in Refs.~\cite{Fukushima2020,Fukushima2026}, and it is better suited to describe ultra-dense (quark) matter in NSs. The singularity in the pressure disappears in the Belinfante EMT. We have therefore analyzed the impact of different prescriptions for the EMT on the nucleon profiles for the density, energy density, and pressures. 

From the profiles obtained in the soliton model, we have constructed an EoS using the direct mapping suggested in Ref.~\cite{Fukushima2020}, and employed in Refs.~\cite{Pradhan2025,Fukushima2026}. We obtained substantial differences between the mechanical EoS and the thermodynamic one. In uniform matter, the two definitions of the pressure should, however, coincide. The obtained differences, therefore, question the method to construct the EoS directly from the profiles of the soliton model.

Since the canonical and Belinfante EMT are pseudo-gauge invariant, the EMT prescription should not impact observables, such as the pressure in ultra-dense matter. We confirm, however, that there is an impact of the EMT prescription on the pressure in ultra-dense matter. This observation may pinpoint a serious caveat in the general approach connecting the energy and pressure profiles in nucleons with the EoS in ultra-dense matter. 

The difference obtained in this paper between the canonical and Belinfante EMT, as well as between the mechanical and thermodynamic pressure, is not expected to appear in ultra-dense matter. Since the direct mapping between the properties of the nucleon and the EoS in ultra-dense matter neglects the surface terms, the differences we obtain may reflect that surface terms are more important than initially expected. An analogy with the liquid-drop model is presented, where the surface terms coincide with the shear force. This analogy presents, however, some limits and does not reconcile the EoS derived from the nucleon model with the thermodynamic one.

In the future, if the ground state of QM can be experimentally studied in nucleon model properties, it will be necessary to extract the contribution of the surface terms in the properties of the Skyrmion. This is possibly the solution to reconcile the mechanical and thermodynamic pressure explored in our current analysis.

\section*{Acknowledgment}

The authors acknowledge discussions with T. Uji and K. Fukushima during the completion of this analysis. B.K.P acknowledges fruitful discussions with D. Mishra and S. Neshatpour at IP2I. The authors acknowledge the support of the CNRS-IN2P3 MAC2 masterproject, the project RELANSE ANR-23-CE31-0027-01 of the French National Research Agency (ANR), the European Union’s Horizon 2020 research and innovation program under grant agreement STRONG–2020-No824093.  This work makes use of \texttt{NumPy} \cite{vanderWalt:2011bqk}, \texttt{SciPy} \cite{Virtanen:2019joe}, \texttt{Matplotlib} \cite{Hunter:2007},  and \texttt{jupyter} \cite{jupyter} software packages.

\bibliographystyle{spphys}    
\bibliography{draft}
\end{document}